\documentclass[runningheads]{llncs}
\usepackage{graphicx}
\usepackage{caption}
\usepackage{subcaption}
\usepackage{cite}

\begin{document}
\title{Prostate Cancer Malignancy Detection and localization from mpMRI using auto-Deep Learning: One Step Closer to Clinical Utilization\thanks{Supported by American Cancer Society.}}
%
%
\author{Weiwei Zong\inst{1}\orcidID{0000-0003-1473-450X} \and
Eric Carver\inst{2} \and
Simeng Zhu\inst{2} \and
Eric Schaff\inst{2} \and
Daniel Chapman\inst{2} \and
Joon Lee\inst{3} \and
Hassan Bagher Ebadian\inst{2} \and
Indrin Chetty\inst{2} \and
Benjamin Movsas \inst{2} \and 
Winston Wen\inst{2} \and
Tarik Alafif\inst{4} \and
Xiangyun Zong \inst{5}}
\authorrunning{Zong et al.}
%
\institute{WeCare.WeTeach, Troy Michigan, USA \and
Henry Ford Health, Detroit Michigan, USA \and
Trinity Health, North Dakota, USA \and
Umm Al-Qura University, Saudi Arabia \and
Shanghai Jiao Tong University Affiliated Sixth People's Hospital, China}

\maketitle              
\begin{abstract}
Automatic diagnosis of malignant prostate cancer patients from mpMRI has been studied heavily in the past years. Model interpretation and domain drift have been the main road blocks for clinical utilization. As an extension from our previous work where we trained a customized convolutional neural network on a public cohort with 201 patients and the cropped 2D patches around the region of interest were used as the input, the cropped 2.5D slices of the prostate glands were used as the input, and the optimal model were searched in the model space using autoKeras. Something different was peripheral zone (PZ) and central gland (CG) were trained and tested separately, the PZ detector and CG detector were demonstrated effectively in highlighting the most suspicious slices out of a sequence, hopefully to greatly ease the workload for the physicians.  

\end{abstract}
%

\section{Introduction}
\label{sec:intro}

In this section, the literature in the field of diagnosis of prostate cancer (PCa) from medical images, the current state-of-the-art research of using deep learning algorithms in automating the diagnosis process and challenges in closing the gap of research and clinical utilization is discussed.

\subsection{Background}

Prostate cancer (PCa) is the most common cancer in men in the United States, and it is the second leading cause of cancer death in these patients (\url{https://www.cancer.org/cancer/prostate-cancer/about/key-statistics.html}). Multiparametric magnetic resonance imaging (mpMRI) for PCa diagnosis has been increased significantly over the past decade. mpMRI sequences have shown promise for the detection and localization of PCa including T2-weighted (T2W), diffusion-weighted imaging (DWI), dynamic contrast-enhanced imaging (DCE) and MR spectroscopy. Combining these MR sequences into a multiparametric format has improved the performance characteristics of PCa detection and localization by evaluating area under curve (AUC) values, sensitivities, specificities, and positive predictive values\cite{delongchamps_multiparametric_2011, dickinson_magnetic_2011}. Prostate biopsy is still considered the golden standard to determine if a suspicious lesion is a benign or a malignant. However, the biopsy is an invasive procedure prone to complications such as hemorrhage, dysuria, and infection. Furthermore, in a small number of cases, the prostate biopsies can fail to establish the diagnosis despite magnetic resonance and transrectal ultrasound-guided approaches4\cite{schouten_why_2017}.

Clinically significant cancer is defined on histopathology as Gleason score 7 (including 3+4) according to PI-RADSTM v2 in order to standardize reporting of mpMRI and correlate imaging findings with pathology results\cite{sadeghi_mp77-02_2018}. Also, PCa is a multifocal disease in up to 87\% of cases; therefore, the ability to distinguish malignant from benign foci within the prostate is crucial for optimal diagnosis and treatment. This has led to an interest in machine learning and computer vision utilizing mpMRI to non-invasively obtain accurate radiologic diagnoses that correlate with their histopathologic variants\cite{meiers_preoperative_2007}. Unfortunately, data scarcity is one of the major challenges in applying deep machine learning algorithms in interpreting mpMRI images due to tightly regulated information acquisition as well as the high cost of MRI acquisition and data labelling from medical experts. In addition, health records cannot be shared without consents due to privacy laws and related healthcare policies and regulations. Recently, efforts have been made\cite{tajbakhsh_convolutional_2016, litjens_survey_2017} to transfer knowledge from publicly available large-scale data consisting of millions or more natural objects or scene images\cite{deng_imagenet:_2009} to medical applications with a small-scale data.

Compared to the recent progress in the application of deep neural networks on 2D images such as chest x-ray\cite{tsai_machine_2019}, mammogram\cite{wu_deep_2020}, the challenges on the classification using prostate MRI\cite{armato_prostatex_2018, Liu_prostate_2017} include relative small patient sample sizes due to high screening cost and much more complex information such as morphological, diffusion or perfusion imaging characteristics. 

\subsection{Previous Work}

Data used in the study was originated from the SPIE-AAPM-NCI Prostate MR Gleason Grade Group Challenge\cite{litjens_computer-aided_2014}, which aimed to develop quantitative mpMRI biomarkers for determination of malignant lesions in PCa patients. PCa patients were previously de-identified by SPIE-AAPM-NCI and The Cancer Imaging Archive (TCIA). 

\begin{figure}[t]
  \centering
   \includegraphics[width=0.8\linewidth]{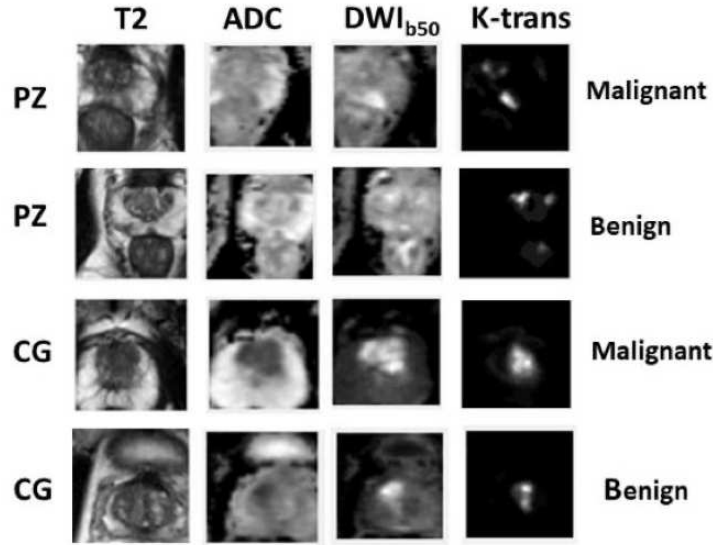}

   \caption{Sample images of the cropped 64x64 pixel rectangle from all four modalities: T2, ADC, DWI, and K-trans. The lesions are malignant in PZ, benign in PZ, malignant in CG, and benign in CG from row 1 to 4 respectively.
}
   \label{fig:fourmodalities}
\end{figure}

Each patient came with four modalities displayed as in Figure~\ref{fig:fourmodalities}. Lesions exhibited hypointense signals in T2 weighted and apparent diffusion coefficient (ADC) map, and hyperintense signals in diffusion weighted imaging with low b values (b = 50). For k-trans, lesions and disease can be confusing, especially in the CG, thus were overlooked in this study.

In our previous work, we found that multi-modality input contributed significantly to accurate classification. In most cases, class activation map (CAM) \cite{zhang_very_2017} helped provide the proof about where the model is looking at when making predictions. The central point of the potential lesion was provided in \cite{litjens_computer-aided_2014}. 

One step closer to clinical utilization by easing the workload of a medical expert is the focus and contribution of this paper. (1) We input the entire prostate gland (PZ and CG separately) rather than the cropped region of interest (ROI). (2) According to Figure~\ref{fig:fourmodalities}, lesion can exhibit very different characteristics when residing in PZ and CG. Another attempt in improving upon our previous work is to train and test using separate models for PZ and CG. (3) To verify the robustness of the trained model, we test it on an independent cohort from our own institute which results in high accuracy. (4) One bonus finding is that, the trained PZ-detector and CG-detector are able to rank the probability of malignancy for each slice and highlight the suspicious slices out of the sequence, despite of the challenges that the testing samples are generated from different scanners with different parameters. This finding served one step closer to clinical utilization.

\section{Data}
\label{sec:Data}
In this section, imaging acquisition parameters and image pre-processing steps are described in detail.

\subsection{Scanning Parameters}
The images were acquired on two different types of Siemens 3T MR scanners, the MAGNETOM Trio and Skyra. T2W images were acquired using a turbo spin echo sequence and had a resolution of around 0.5 mm in plane and a slice thickness of 3.6 mm. The DWI series were acquired with a single-shot echo planar imaging sequence with a resolution of 2 mm in-plane and 3.6 mm slice thickness and with diffusion-encoding gradients in three directions. Three b-values were acquired (50, 400, and 800 s/mm2), and subsequently, an ADC map was calculated by the scanner software.

To test cross-institutional generalization capability of our model, an independent cohort (test cohort 2) consisting of 40 patients was collected for testing from our own institution with the IRB-approval. An ultrasound-guided needle biopsy was performed to confirm the diagnosis. Two image modalities were acquired for each patient using a 3.0 T MR scanner (Ingenia, Philips Medical System, Best, the Netherlands) using small field of view as follows: T2W acquired with Fast-Spin-Echo (TE/TR: 4389/110ms, Flip Angle: 90$^{\circ}$ with image resolution of 0.42×0.42×2.4mm3) and DWI with two b values (0 and 1,000 s/mm2). The voxel-wise ADC map was constructed using these two DWIs.

\subsection{Data Pre-Processing}

\begin{figure*}
  \centering
  \includegraphics[width=0.8\linewidth]{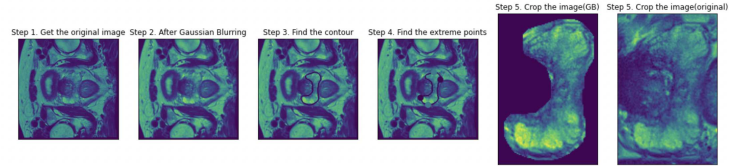}
  \caption{PZ cropping steps on one slice of the sequence. CG cropping follows similar steps.}
  \label{fig:crop1slice}
\end{figure*}

All images were registered to T2-axial image sets and resampled to 1 mm isotropic resolution.
N4 bias-field correction was applied to T2 and ADC for intensity uniformity correction. Then sub-regions such as PZ or CG were cropped from the sequence based on the contouring performed by our team of medical experts.  Pre-processing and cropping on one slice were shown in Figure~\ref{fig:crop1slice}. Gaussian blurring was applied to increase the contrast. Four extreme points were located based on the annotated contour. With a margin of 5 pixels, a surrounding rectangle was cropped. 

\begin{figure*}
  \centering
  \includegraphics[width=0.8\linewidth]{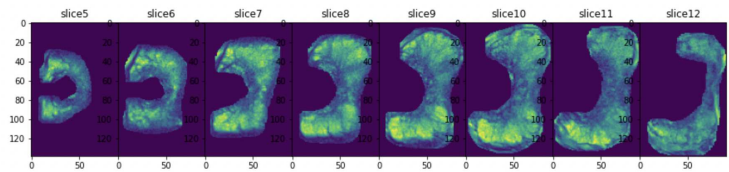}
  \caption{PZ cropping results on one sequence consisting of 8 slices.}
  \label{fig:cropseq}
\end{figure*}

\begin{figure}
  \centering
  \begin{subfigure}[b]{0.45\textwidth}
  \centering
    \includegraphics[width=0.85\linewidth]{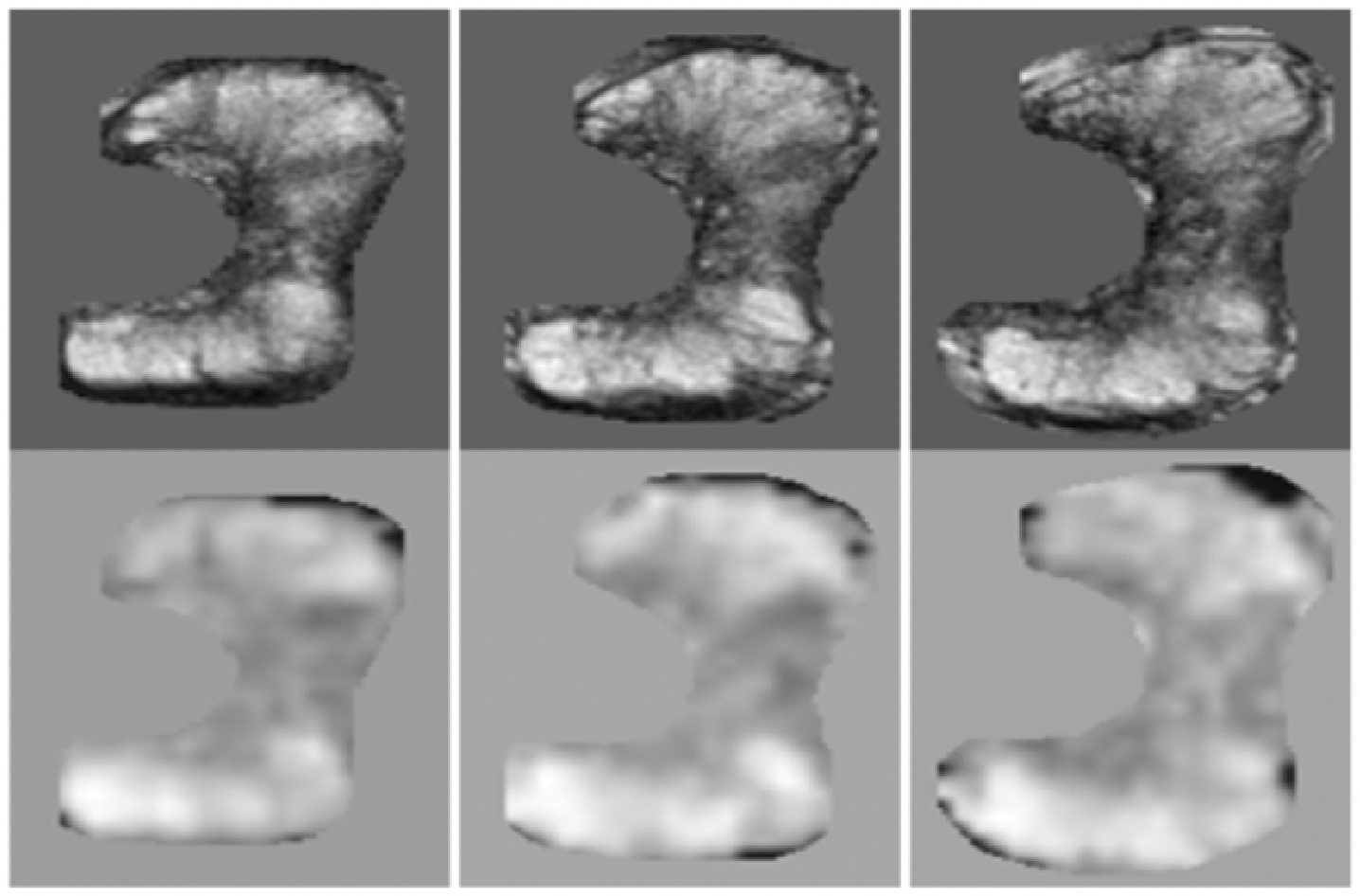}
    \caption{PZ with Gleason Grade Group 3.}
    \label{fig:crop3-pz}
  \end{subfigure}
  \begin{subfigure}[b]{0.45\textwidth}
  \centering
    \includegraphics[width=0.85\linewidth]{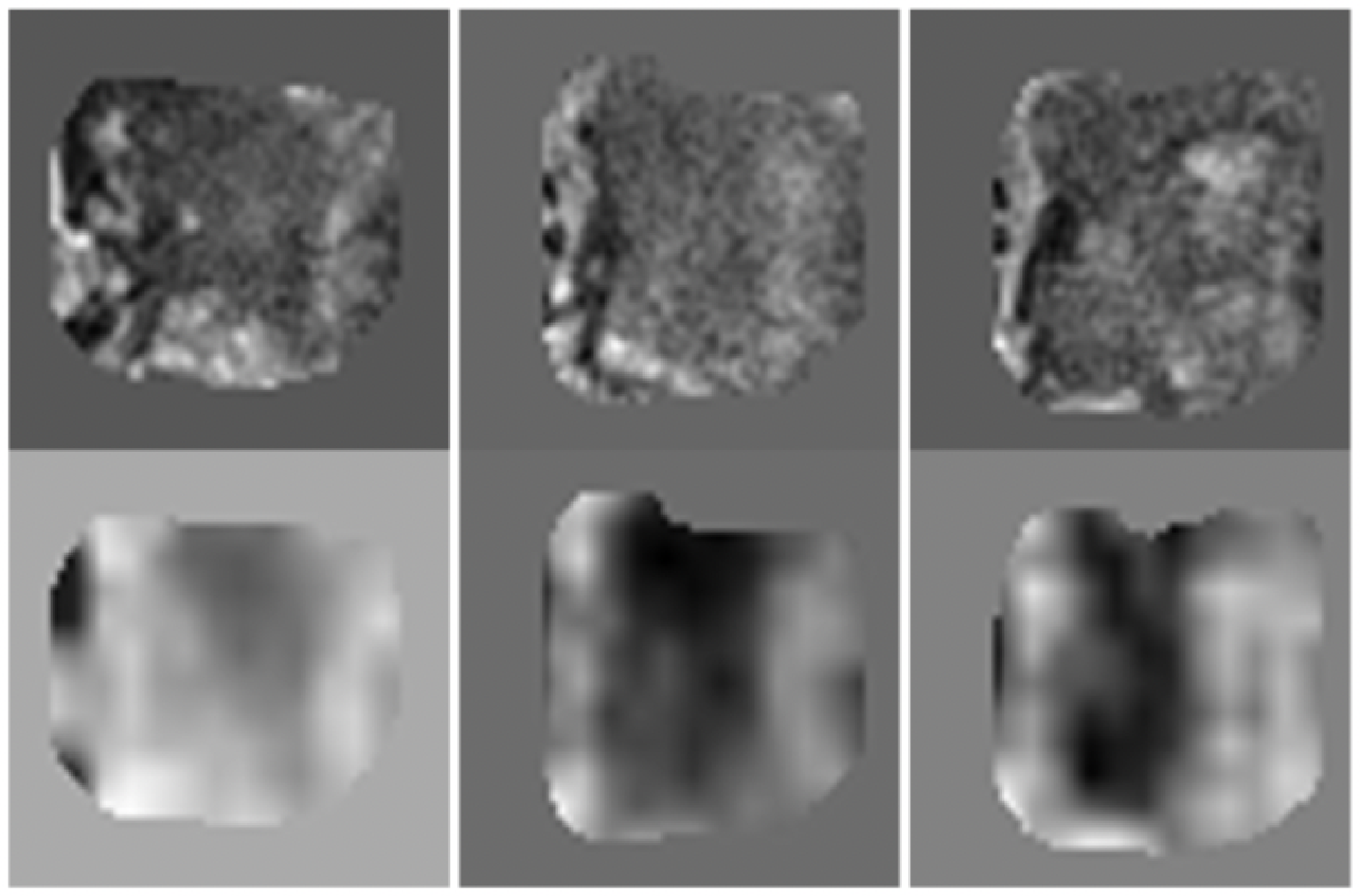}
    \caption{CG with Gleason Grade Group 2.}
    \label{fig:crop3-cg}
  \end{subfigure}
  \caption{Example of cropped sub-regions with three consecutive slices.}
  \label{fig:crop3}
\end{figure}

During training and testing, the sequence representing one patient were taken into consideration. Therefore, the extreme points were searched through the sequence to not to miss any part of the PZ or CG, as shown in Figure~\ref{fig:cropseq} and Figure~\ref{fig:crop3}. 

\section{Method and Results}
\label{sec:Method}

\subsection{Auto-Keras}
Auto deep learning models for medical image analysis have not been studied much. To fully take advantage of this technique, implementation details were illustrated as follows.

To increase the sample size, and to leverage the 3D information between two consecutive slices, the 2-channel input was augmented to include T2-ADC pair, consecutive T2 pairs and consecutive ADC pairs. Performance of using just T2-ADC pair and the mixed pairs were compared for both PZ and CG detectors. 

The number of benign lesions outnumbered malignant lesions and re-weighting was used accordingly. After shuffling the data, third of the data were randomly selected for validation. Early stopping with a patience value of $10$ and optimization goal of area-under-the-curve (AUC) for validation set was used to prevent overfitting. Bayesian tuner was used as the searching strategy and maximum searching epochs were set as $20$. Auto-augmentation included a search in the augmentation space with operations such as random flip, translation, and contrast, etc. 

\begin{figure}
  \centering
  \begin{subfigure}[b]{0.40\textwidth}
  \centering
    \includegraphics[width=0.85\linewidth]{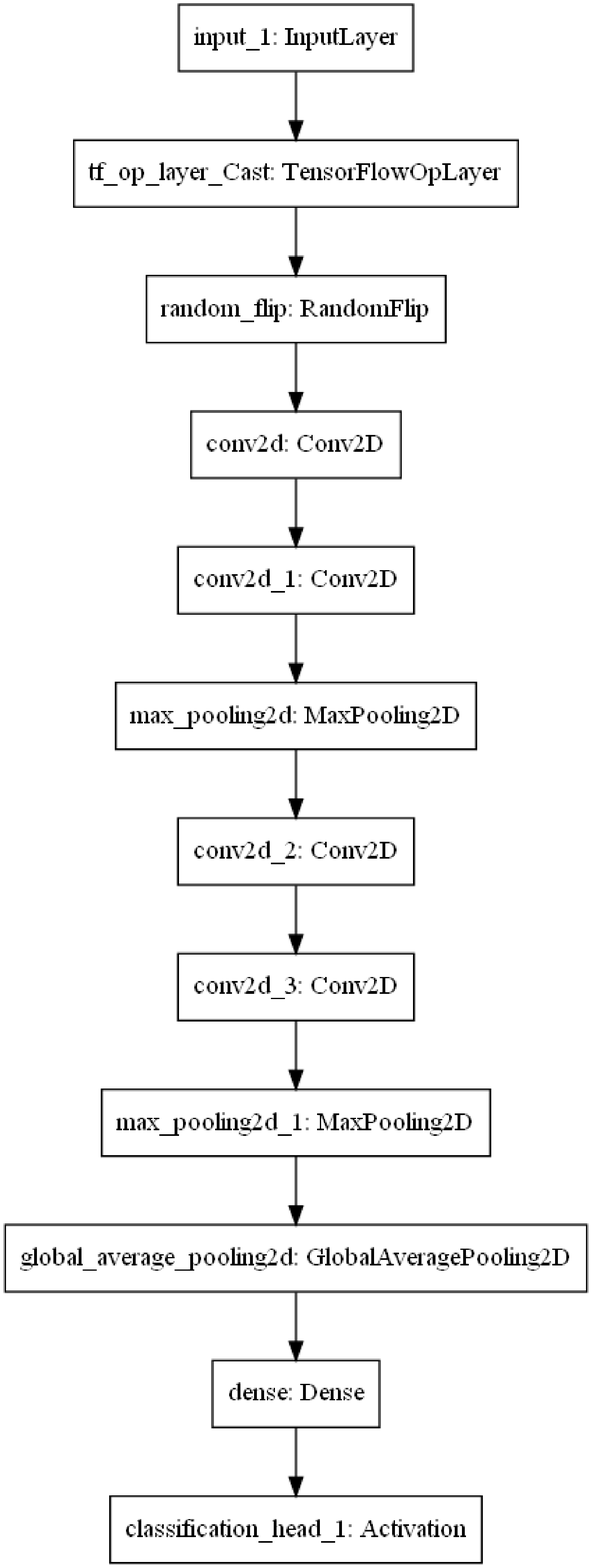}
    \caption{PZ-detector architecture.}
    \label{fig:autonet-pz}
  \end{subfigure}
  \begin{subfigure}[b]{0.40\textwidth}
  \centering
    \includegraphics[width=0.85\linewidth]{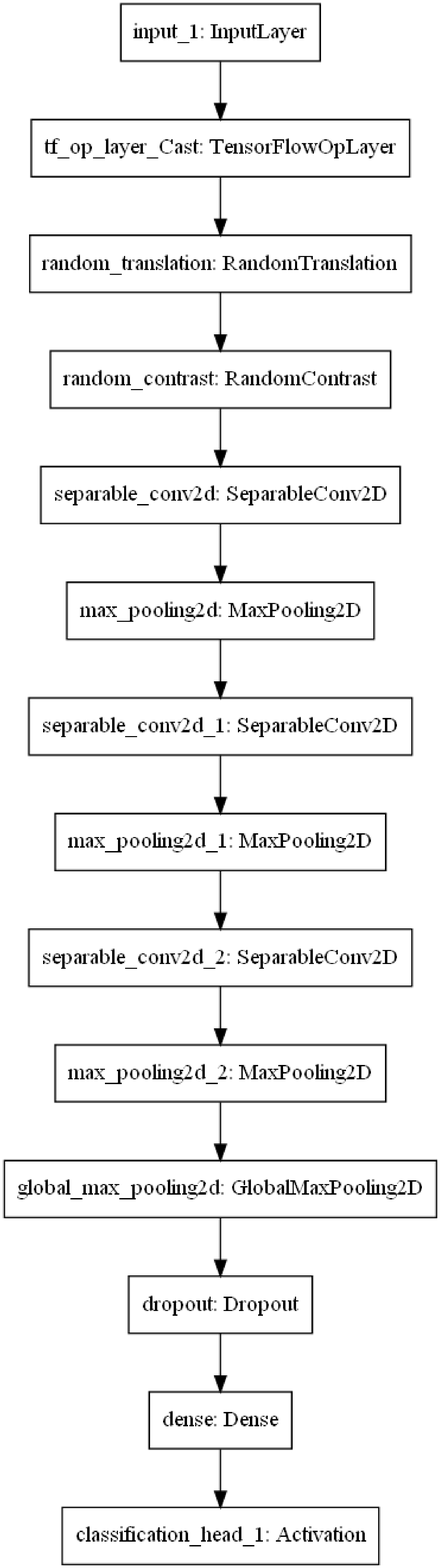}
    \caption{CG-detector architecture.}
    \label{fig:autonet-cg}
  \end{subfigure}
  \caption{Customized models to detect lesions in PZ (left) and CG (right) respectively.}
  \label{fig:autonet}
\end{figure}

The learned optimal models to detect lesions in PZ ~\ref{fig:autonet-pz} and CG ~\ref{fig:autonet-cg} were displayed in Figure~\ref{fig:autonet}. Compared to PZ-detector, data augmentation and network architecture were found more complicated to recognize malignant lesions in CG. However, the best validation AUC of 0.94 was achieved for CG-detector when only T2-ADC pairs were used as the input. According to Figure~\ref{fig:fourmodalities}, lesions in CG are visually difficult to detect from single source of modality, which might have explained the preference of T2-ADC pair as the input. The best validation AUC for PZ-detector, on the other hand, was 0.90, when the input was a mixture type of pairs.

\begin{figure}
  \centering
  \begin{subfigure}[b]{0.45\textwidth}
  \centering
    \includegraphics[width=0.85\linewidth]{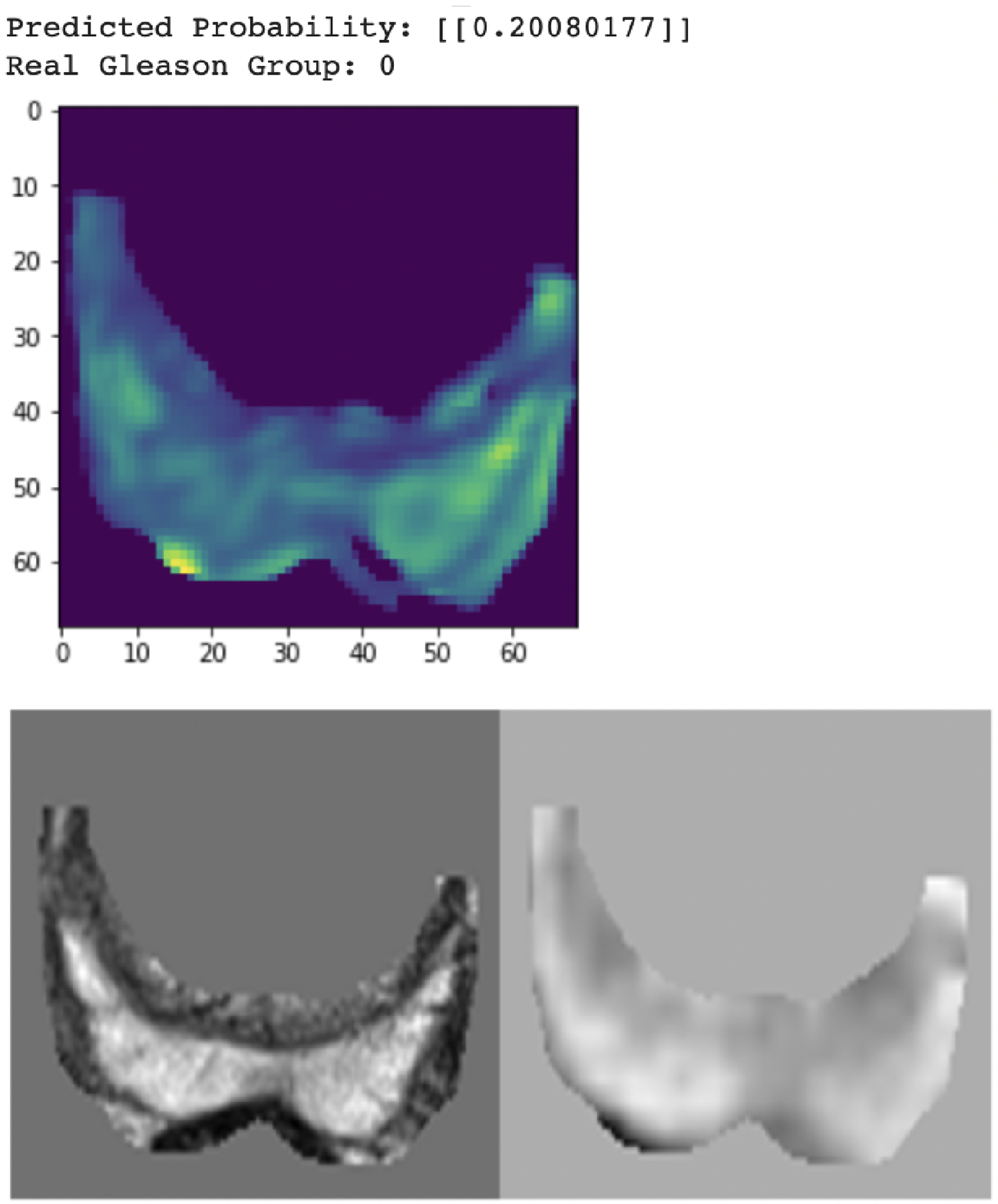}
    \caption{Grad-CAM for one T2-ADC pair of PZ with GGG 0.}
    \label{fig:traincam-pz}
  \end{subfigure}
  \begin{subfigure}[b]{0.45\textwidth}
  \centering
    \includegraphics[width=0.85\linewidth]{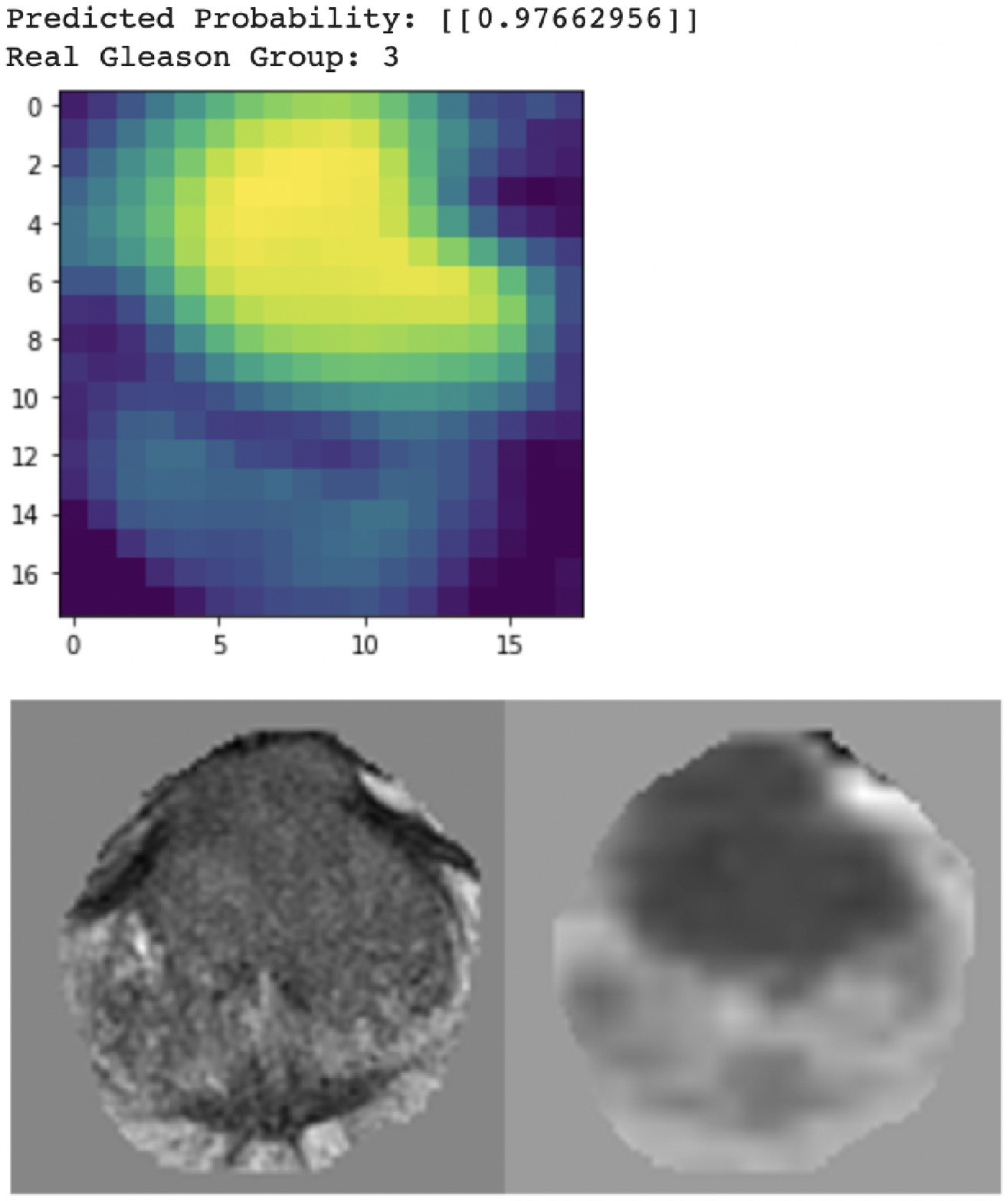}
    \caption{Grad-CAM for one T2-ADC pair of CG with GGG 3.}
    \label{fig:traincam-cg}
  \end{subfigure}
  \caption{Grad-cam for PZ (left) and CG (right) respectively, showing the attention of the model aligned well with the lesion if malignant.}
  \label{fig:traincam}
\end{figure}

In Figure~\ref{fig:traincam}, gradient class activation map (Grad-CAM) was used to visualize the learned PZ-detector ~\ref{fig:traincam-pz} and CG-detector ~\ref{fig:traincam-cg}. The prediction probability was the value of the output node. It can be observed from the two cases that, lesions aligned well with the hyper-intense signal in the Grad-CAM if predicted malignant, otherwise neither hyper- nor hypo-intense signal was visible.

\subsection{Challenges of Domain Drift}
Testing data cohort was collected from our institute and posed several challenges during the inference stage. 

(1) Contours of PZ or CG were not available on contrary to the cases in the training cohort, but more of a realistic setting. 

(2) The ADC maps were calculated from DWI $b = 0 and b = 1, 000$. They are visually slightly different from the ADC maps in the training cohort. 

(3) Number of slices were varied in the testing cohort. 

\subsection{Solutions}
\begin{figure*}
  \centering
  \includegraphics[width=0.8\linewidth]{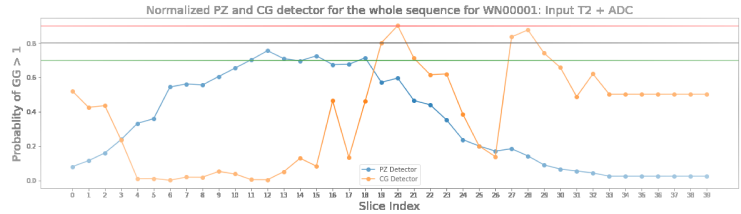}
  \caption{Predictions made on the sequence of Patient $WN00001$ from the independent testing cohort by PZ-detector (blue) and CG-detector (orange), respectively.}
  \label{fig:wn001}
\end{figure*}

(1) We nevertheless tested PZ-detector and CG-detector on the testing cohort, where the input is the T2-ADC pair and whole prostate gland was used as the input since sub-region contour was not available. The PZ-detector was able to accurately highlight slices with suspicious malignant lesions. While the CG-detector was able to detect lesions in CG most of the time but were found prone to the false positive traps as shown in Figure~\ref{fig:wn001}, such as the marginal Slice $28, 29$.

\begin{figure*}
  \centering
  \includegraphics[width=0.8\linewidth]{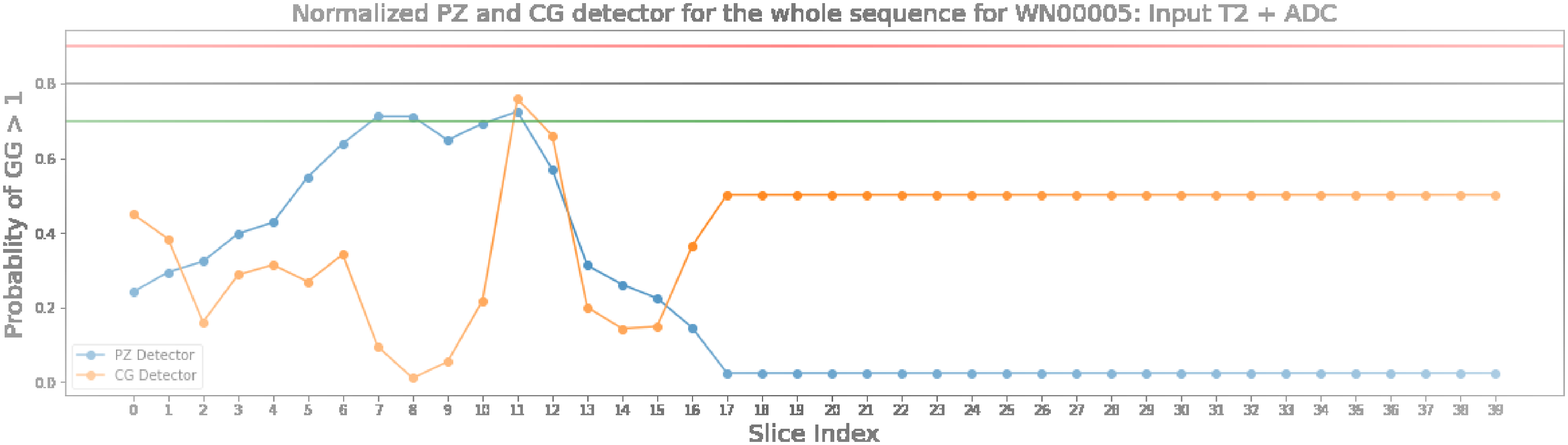}
  \caption{Predictions made on the sequence of Patient $WN00005$ from the independent testing cohort by PZ-detector (blue) and CG-detector (orange), respectively.}
  \label{fig:wn005}
\end{figure*}

Figure~\ref{fig:wn005} was one of the successful cases where both CG-detector and PZ-detector were able to pick the suspicious slice and the Grad-CAM showed the location of suspicious lesion on that slice which aligned well with the lesion.

\begin{figure}[t]
  \centering
   \includegraphics[width=0.5\linewidth]{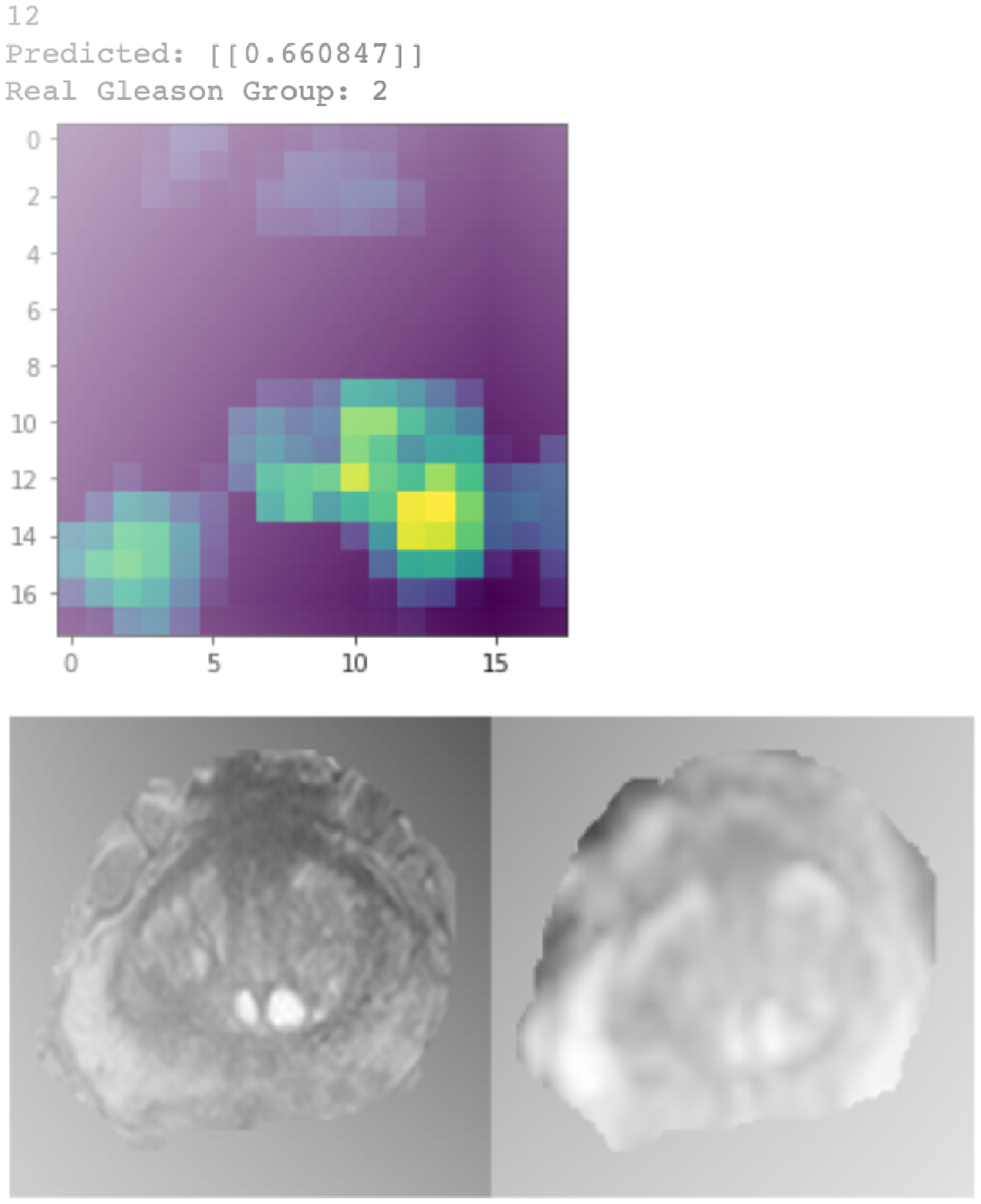}

   \caption{Grad-CAM from the CG-detector indicating PZ-detector and CG-detector were both sensitive to Slice $12$ for Patient $WN00005$, which showed hyper-intense signal in both PZ and CG.
}
   \label{fig:wn005cam}
\end{figure}

(2) The trained models were verified robust to the domain drift issues caused by ADC map calculated from DWIs with different b values.

(3) By making prediction for each slice and connecting the slices into the sequence, the PZ-detector and CG-detector were able to work together to highlight the suspicious slice out of the sequence.

\section{Conclusions}

This work extended our previous research on using deep learning models to read from mpMRI and suggest diagnosis for lesion malignancy. The purpose was one step closer to clinical utilization by means of eliminating the manual efforts of lesion localization, utilizing automatic deep learning framework to search for the optimal augmentation strategy, network architecture and parameters, and finally make prediction for the sequence to find the most suspicious slice and localize the lesion on that slice using the saliency map. The code has been made public and ready to be deployed for anyone who is interested.

\bibliographystyle{splncs04}
\bibliography{pcabib}

\end{document}